\documentstyle[12pt,rotating,epsfig,amsmath,here,amssymb,times,hhline]{article}
\newlength{\dinwidth}
\newlength{\dinmargin}
\title{The VLQ Calorimeter of H1 at HERA: \\ A Highly Compact  
       Device for Measurements of Electrons and Photons under 
       Very Small Scattering Angles}

\begin{document} 

\maketitle

\begin{center}

 A.~Stellberger$^{1}$,            %HDB2
 J.~Ferencei$^{2}$,               %HDB2-PD                  Ferencei
 F.~Kriv\'{a}\v{n}$^{2}$,         %KOSI-TP                  Krivan
 K.~Meier$^{1}$,                  %HDB2-PD                  Meier
 O.~Niedermaier$^{1}$,            %HDB2-ST
 O.~Nix$^{1}$,                    %HDB2-PD
 K.~Schmitt$^{1}$,                %HDB2
 J.~\v{S}palek$^{2}$,             %KOSI-TP                  Spalek
 J.~Stiewe$^{1}$,                 %HDB2-PD     1/93         Stiewe
 and
 M.~Weber$^{1}$.                  %DESY-PD                  Weber2

\date{}

\vspace{1cm}
\noindent

 $\:^1$ Kirchhoff - Institut f\"ur Physik der Universit\"at Heidelberg,
        Heidelberg, Germany    \\
 $\:^2$ Institute of Experimental Physics, Slovak Academy of
        Sciences, Ko\v{s}ice, Slovak Republic   \\
	
\end{center}

\vspace{1cm}

%\abstract{
\begin{abstract}
In 1998, the detector H1 at HERA has been equipped with a small
backward spectrometer, the Very Low Q$^2$ (VLQ) spectrometer comprising
a silicon tracker, a tungsten - scintillator sandwich calorimeter,
and a Time-of-Flight system. The spectrometer was designed to measure
electrons scattered under very low angles, equivalent to very low
squared four -
momentum transfers $Q^2$, and high energy photons with good energy
and spatial resolution. The VLQ was in operation during the 1999 and 2000 
run periods. This paper describes the design and construction of the VLQ 
calorimeter, a compact device with a fourfold projective energy read-out, 
and its performance during test runs and in the experiment.
\end{abstract}

\vspace{2cm}

\section{Introduction}

\subsection{Physics Motivation}
The storage ring HERA at DESY where electrons or positrons of 27.5 GeV 
collide with protons of 920 GeV is devoted to the investigation of deep 
inelastic electron - proton scattering (DIS) and photoproduction 
processes. The detector H1 
\cite{H1det} at HERA is equally well equipped for measuring electrons 
scattered in deep inelastic, i.e. high $Q^2$, reactions, by virtue of 
its Liquid Argon (LAr) \cite{H1det} and SpaCal \cite{SpaCal} calorimeters, 
and electrons which initiated photoproduction, i.e. $Q^2 \approx 0$,
reactions with the help of its low angle (w.r.t. the electron beam 
direction) electron spectrometers, called the ``electron taggers''.

\begin{figure}[h]
  \center
   \includegraphics[width=13cm]{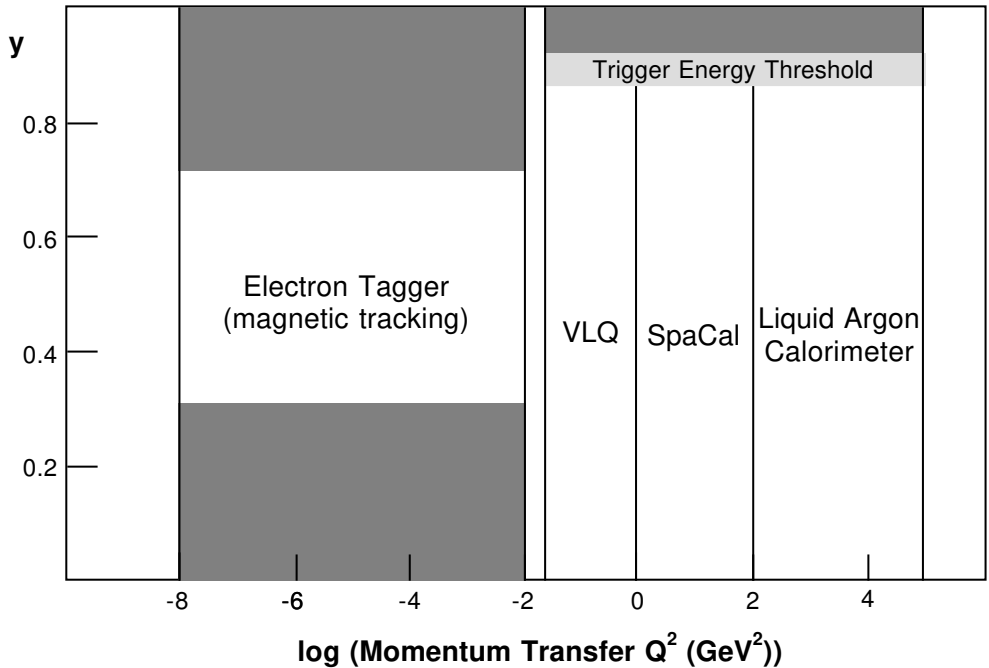}
  \caption{The acceptance region of H1 in the $y$ - $Q^2$ plane.}
    \label{fig:kinpl}
\end{figure}

However, the available range in $Q^2$ which is linked to the electron's
scattering angle is limited by the inner radius of the SpaCal which
corresponds to a minimal $Q^2$ of roughly 0.6 GeV$^2$. Therefore, a
large area in phase space cannot be accessed by H1 without the help of a
dedicated device, see Fig. \ref{fig:kinpl} which shows 
the acceptance range of H1 in a $Q^2 - y$ map. The variable
$y$ is, for vanishing $Q^2$, defined by $~y=1-E'/E~$ where $E$ is the
energy of the incoming, $E'$ that of the scattered electron. Therefore,
$y~=~0$ corresponds to vanishing, $y~=~1$ to maximal photon energy.
The region below the SpaCal - covered
$Q^2$ range is of special interest because here the 
DIS and photoproduction regimes merge: The ``transition region''. A
measurement of the structure function $F_2(W^2,Q^2)$, or equivalently
of the total photo-absorption cross section 
$$  \sigma(\gamma^* p) = 4~\pi^2~\alpha/Q^2~F_2(W^2,Q^2), $$
in this region is expected to shed light on the
interplay between soft and hard processes in electron - proton
scattering. The VLQ extends the kinematically accessible range down 
to values of $Q^2 = 0.03$.

Furthermore, a measurement of vector meson production in this
regime, especially a comparison between cross section behaviours for
the production of light ($\rho^0$, $\omega$, $\phi$) and heavy (J/$\psi$,
$\Upsilon$) mesons, in the transition region will help to improve 
the understanding of strong interaction mechanisms \cite{Proposal}.

The VLQ calorimeter, which does not distinguish between high energy
electrons and photons, can also serve to measure photons from meson
decays, e.g. from $\pi^0$ or $\eta$ mesons, or from heavier particles
which decay through these intermediate states. A field which gained
new interest is meson production via ``Odderon'' exchange, where the
Odderon is the $P = C = -1$ partner of the Pomeron. Odderon - mediated
reactions are expected to lead to high energy single meson production 
in the ``photon hemisphere'', that is in the electron beam direction
\cite{DoNa}.
The VLQ is an ideal target for photons from these meson decays 
\cite{oddpom}.

\subsection{Detector Design: Constraints and Concept}

The VLQ spectrometer consists of two independent modules, each
comprising a tracker system and an electromagnetic calorimeter, 
followed by a time-of-flight (ToF) system which has 
the purpose to recognize background particles travelling along with the
proton beam. The phase space region to be covered dictates the position
of the VLQ to be downstream the electron beam and close to the beam
pipe. The space available to install a new detecting device there is
strongly limited, enforcing a very tight structure, and especially a
highly compact calorimeter. The solution chosen is a system of two
identical spectrometer modules, one above and one below the beam line,
fixed to a moving mechanics so that the modules can be moved close to
the beam, or retracted, independently. The tracker part and the ToF 
system connected to the VLQ spectrometer are described in \cite{Hurling}.

The volume available for either VLQ calorimeter module is a box of 
16.3 cm length and transverse dimensions of 15.0 x 18.0 cm$^2$. The
calorimeter operates in an energy range up to the electron beam
energy, i.e. roughly 30 GeV. The energy resolution aimed at in this
regime is 3 - 4 \%. A material capable to absorb electromagnetic
showers of this energy within the given distance is tungsten, with
Z = 72 and a density of 19.3 g/cm$^3$. A sandwich calorimeter type was
chosen in a sophisticated layout which allows a ``projective 
energy read-out'',
i.e. a fourfold projection of the shower profile can be measured along
the shower detector: A passive 
 tungsten~\footnote{For financial and
 machining reasons, the material used consists of 95 \% tungsten and a
 5 \% admixture of nickel and copper.} 
plate of 2.5 mm thickness is 
followed by an active scintillator layer of 3 mm thickness, structured 
in an array of 24 vertical scintillator bars of 5 mm width each; 
this double layer is continued by another tungsten plate and another 
scintillator layer, arranged in 18 horizontal scintillator bars of the 
same width. In total there are 23 tungsten and 24 scintillator layers. 
The scintillation light which is guided by total reflection is
read out at either end of each bar, and coupled 
into a wavelength shifter which guides the light towards 
photodiode detectors.

\section{Calorimeter Design and Construction}

\subsection{Design Parameters and Optimization}

\begin{figure}[h]
  \center
     \includegraphics[width=13cm]{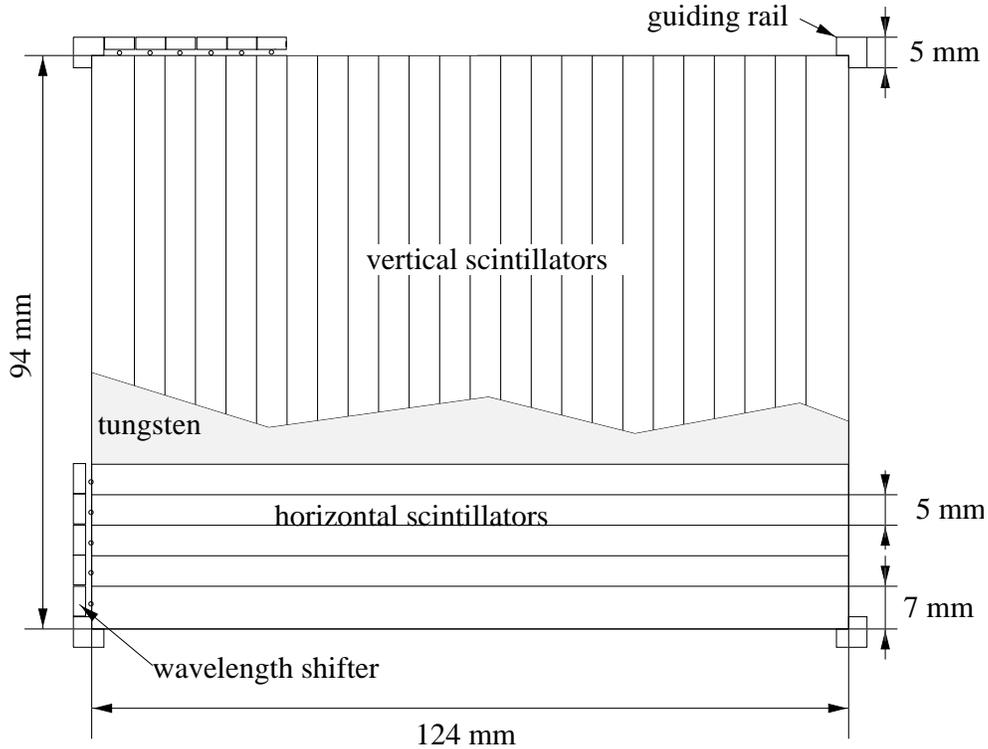}
  \caption{Front view of the calorimeter structure.}
    \label{fig:calofront}
\end{figure}

\begin{figure}[h]
  \center
     \includegraphics[width=13cm]{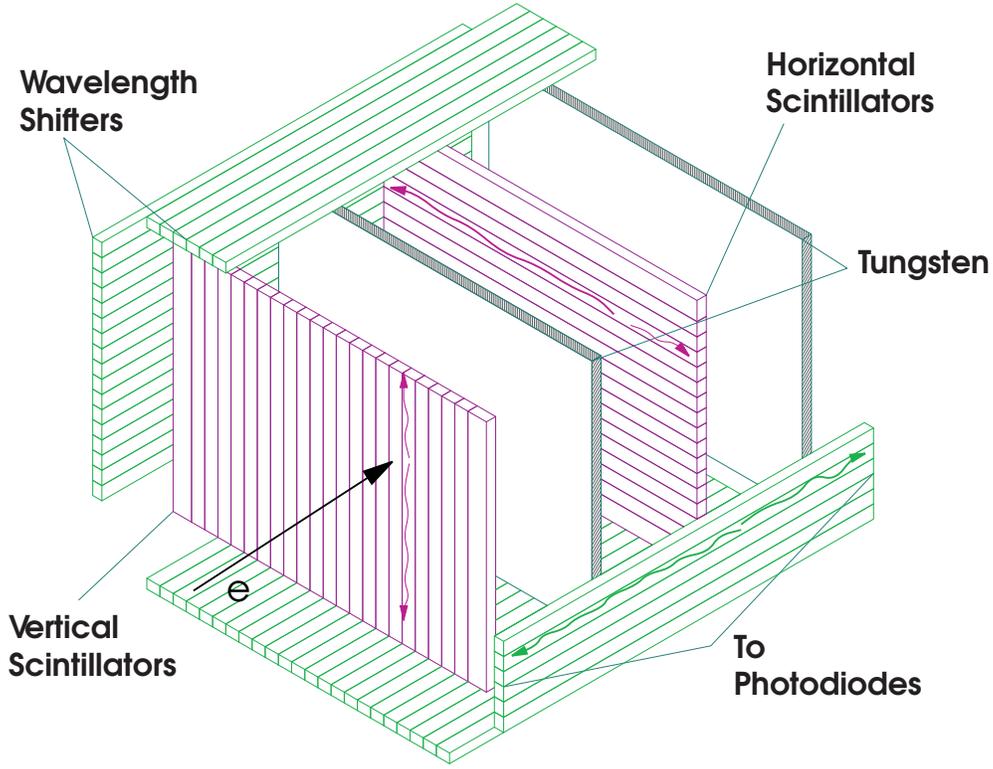}
  \caption{``Exploded view'' of a calorimeter module.}
    \label{fig:explov}
\end{figure}

Detection devices positioned very close to the beam pipe are exposed to
high doses of synchrotron radiation and background generated by the
proton beam.
In order to protect the VLQ from background radiation, a moving
mechanism retracts the spectrometer from the beam pipe
during the filling phase. This condition renders a full $2 \pi$ 
coverage in azimuth more difficult and leaves a two module solution: 
One module above, and another
identical one below the beam pipe, fixed to the retraction mechanism which
moves the modules to well defined measuring positions. The parameters
of the calorimeter described below always refer to a pair of modules.
The structure of a calorimeter module is sketched
in Fig.~\ref{fig:calofront} , an ``exploded view'' is shown 
in Fig.~\ref{fig:explov}.

The scintillation material chosen is Bicron BC-408 with a
wavelength of maximum emission of 425 nm, a decay time 
of the scintillation light of 2.1 ns,
and a refractive index of 1.58~\cite{BCsheet1}.

Given the available space and the absorber and scintillator materials, the
internal structure of the calorimeter module has to be optimized. The total
length being fixed, the parameters to be varied are number and thickness 
of tungsten vs. scintillator plates. The optimization procedure was carried
out using the simulation package GEANT \cite{geant}. The optimization 
criterion is derived from the behaviour of the relative energy resolution
$\sigma(E)/E$ which is conventionally expanded in powers of $1/\sqrt{E}$:
\begin{equation}
       \sigma(E)/E = S/\sqrt{E} \oplus C \oplus N/E .
\end{equation}
Here, $S/\sqrt{E}$ is the so called ``sampling term'' governed by 
Poisson statistics
of shower sampling. $C$ is called the ``constant term'' saying that the
absolute energy resolution $\sigma(E)$ worsens with increasing energy. 
It is mainly due to incomplete shower containment and strongly dependent
on the calorimeter size. $N/E$ is the noise term which is not a property of
the calorimeter body itself but given by the readout system. It has 
the same influence on the absolute resolution at any energy, i.e. it is
independent of energy.

Optimization of the mechanical layout of the calorimeter can be decoupled
from the electronic read-out \cite{Achim2}. To find the best solution, a
series of simulations was performed where the tungsten and scintillator 
layer thicknesses were varied \cite{Achim2}.  The electron impact
energies were 5, 15 and 25 GeV. A curve corresponding to eq. 1 was fit
to the simulated energy distribution without the noise term which has
not been simulated. The
optimal solution, a tungsten layer thickness of 2.5 mm and a scintillator
layer thickness of 3 mm, resulted in the following expression for the 
energy resolution:

\begin{equation}            
  \frac{\sigma(E)}{E} = \sqrt{\left(\frac{12.9 \%}{\sqrt{E/\text{GeV}}}\right)^2
                        \oplus (3.2 \%)^2} .
\end{equation} 
A parameter for the noise term could not be derived from the
simulation but was obtained from an independent 
investigation of the read-out chip, 
see \cite{Achim2, VLQshort} and below (Section 2.4). 

\newpage

An equivalent investigation was made in order to determine the optimal
width of the scintillator strips. A good spatial
resolution is clearly in favour of a small width,
but implies a larger number of read-out channels 
and a smaller light yield per scintillator, with increased relative noise
contributions. The compromise found was a width of 5 mm resulting in a (one 
dimensional) spatial resolution of 1.4 mm at 5 GeV. The resolution improves
to below 1 mm for higher energies due to reduced shower fluctuations.

\begin{figure}[h]
  \begin{center}
     \includegraphics[width=8cm]{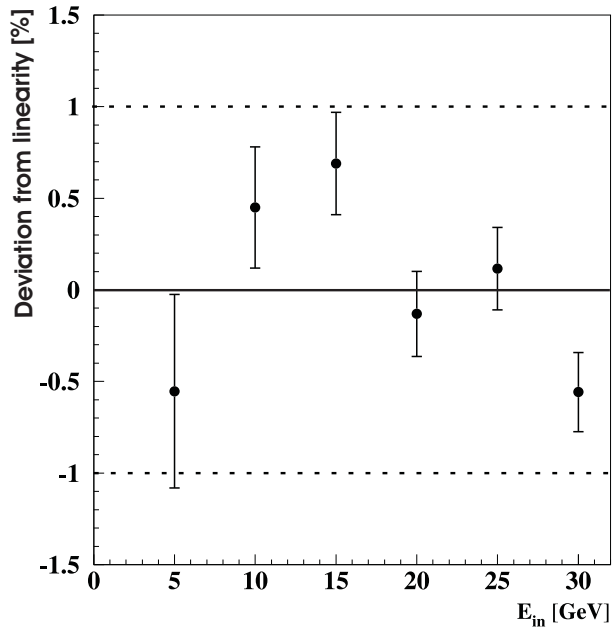}
\begin{quote}
  \caption{Deviation of calorimeter response from linearity in the energy
           range from 5 to 30 GeV. Result of simulation.}
    \label{fig:lineasim}
    \end{quote}
    \end{center}
\end{figure}

Another important property of any calorimeter is the relation between
input energy and measured response. Deviations 
from linearity lead to systematic uncertainties and deteriorate the
effective
energy resolution. The simulation study on the optimized VLQ calorimeter
shows deviations from linearity below one percent 
which are a consequence of the compactness of the calorimeter body
and corresponding leakage effects at high energies, 
see Fig. \ref{fig:lineasim}.

Following the simulation results, the calorimeter in its structure as 
described above represents an ``electromagnetic depth'' of 16.7 radiation 
lengths and has a Moli\`ere radius (the radius of a cylinder containing 
on the average 90 \% of the shower energy) of 1.25 cm.

\subsection{Mechanical Layout}

The two modules of the calorimeter need to be retracted during
the beam filling phase in order to protect the silicon tracking device
from synchrotron and background radiation. With the help of the moving
mechanics, the radiation load of the VLQ tracking system can be reduced
by a factor of more than 20.
The moving apparatus delivers a (vertical) 
position information which is read out and continuously stored
after every move-in / move-out
procedure. Its accuracy is of the same order as that of the tracker, 
i.e. a few $\mu$m.

\begin{figure}[h]
  \center
     \includegraphics[width=8cm]{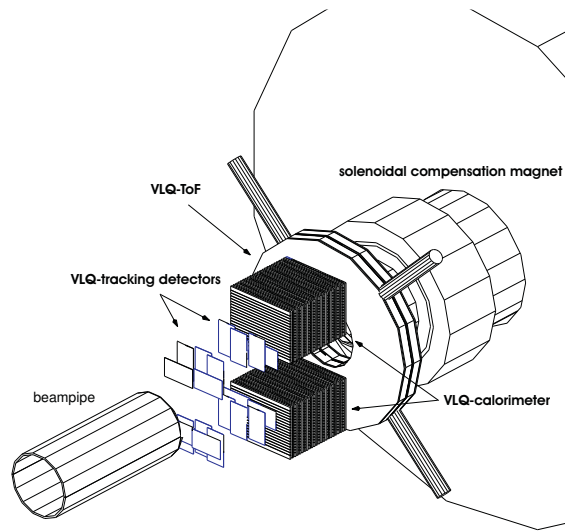}
  \caption{Sketch of the mechanical layout of the VLQ (tracker,
           calorimeter, ToF).}
    \label{fig:VLQlayout}
\end{figure}

The VLQ calorimeter modules, including the electronic read-out systems, 
are accomodated in a solid brass housing each, with wall thicknesses 
of 8 mm (side walls), 5 mm (front and back walls), and 2 mm (bottom wall,
i.e. the wall closest to the beam pipe). Thus, sufficient protection 
against synchrotron radiation is provided. 
Inside a module, another 10 mm thick brass plate is mounted
directly above the active volume which accomodates the cooling water tubing
system: The water serves to cool an air stream which continually flows
around the electronics boards.
The tracker modules are fixed to the calorimeter modules, as 
sketched in Fig. \ref{fig:VLQlayout}.

\begin{figure}[h]
  \center
     \includegraphics[width=8cm]{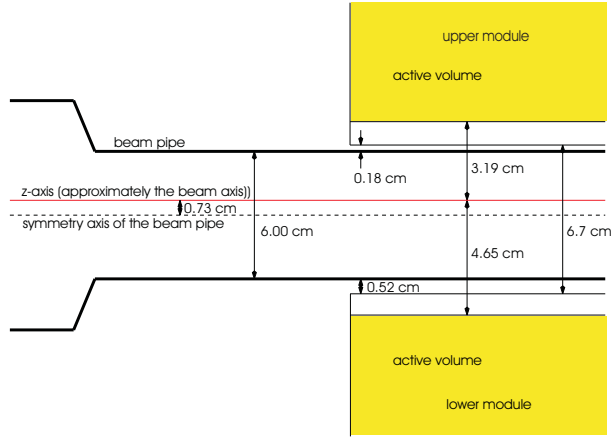}
  \caption{Sketch of the beam pipe shape in front of the VLQ
           spectrometer.}
    \label{fig:beampipe}
\end{figure}

The VLQ spectrometer is situated as close to the beam line as possible
in order to catch electrons and photons scattered or produced at very
small angles. This requires a modification to the beam pipe because 
even a thin but straight cylindrical pipe would force particles to 
traverse a large amount of material. The shape of the beam pipe with
its changing diameter is shown in Fig. \ref{fig:beampipe}. 
The pipe is made of aluminium, and the material in front of the VLQ 
corresponds to less than one radiation length.

\newpage

\subsection{Light Transport and Collection}

After the considerations sketched above, the structure and granularity of
the scintillation and light transport system are fixed: The calorimeter
has 24 scintillator layers, 12 vertically and 12 horizontally structured 
ones. Each vertically structured layer consists of 24 bars, each 
horizontally structured one consists of 18 bars. An exploded view of
the calorimeter structure is displayed in Fig. \ref{fig:explov} . 
The active, scintillating layers are interleaved with 23 passive, tungsten 
layers. The very first and very last passive layer are represented by the 
front and back brass housing walls.

Each scintillator strip is read out at either side
by a wavelength shifter (WLS)
bar~\cite{BCsheet2} which collects and sums the light from 24 strips: In 
total, there are 2 x (18 + 24) = 84 WLS bars per module. Each WLS bar is
read out by a photodiode at either end so that 168 photodiodes are needed
per module. Each surface plane of a calorimeter module is read out
in this manner, so that four independent transverse shower projections are
available.
 
The wavelength shifters are located close to each other with a distance of 
only 0.2 mm. Consequently, the use of a single photodiode with its
own housing for an individual WLS is not possible. Instead, especially
manufactured photodiode arrays which match the dimensions of the WLS
arrays were employed
 \footnote{Silicon PIN diode arrays manufactured by Hamamatsu Photonics}. 
There are two kinds of arrays, one with 18 and one with 24 diodes, 
each implanted in a common p - substrate,
corresponding to the transverse structure of the
calorimeter with its two different sets of read-out channels. The size of 
the active area of a single diode is 4.2 x 3.4 mm$^2$. The distance between
the active areas is 0.8 mm. The diode arrays are protected by a glass
surface of 0.5 mm thickness.

The determination of shower energy and shower position is based on the
evaluation of the transverse shower profiles. Any effect which disturbs
the primary ``physical'' profile has to be avoided or at least minimized.
One such effect ist optical cross talk between adjacent scintillator 
strips: Scintillation light induced in one channel reaches neighbouring
ones and thus distorts the primary profile. To avoid this, each scintillator
strip was wrapped in individually prepared and cut white paper.

Each scintillator strip matches with either end its WLS bar. Scintillator
and WLS have the same refractive index (n = 1.58) so that an intimate contact
would be desirable to minimize light losses. Since this contact could only
be established manually by gluing which inevitably leads to non-uniform
connections, scintillator and WLS were separated by a 0.2 mm wide gap
which is maintained by a Nylon thread. The WLS strips guide the light
directly onto the photodiode surfaces.

\begin{figure}[h]
  \center
     \includegraphics[width=8cm]{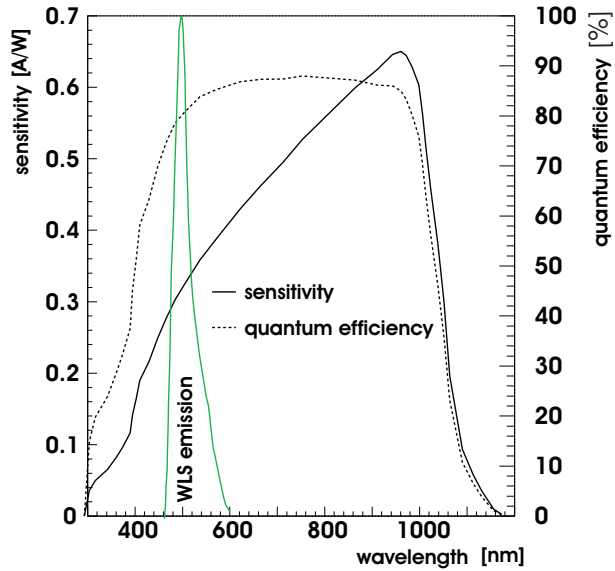}
  \caption{Response curves of the Hamamatsu PIN photodiode.}
    \label{fig:photodio}
\end{figure}

Scintillator and WLS materials are selected to match the sensitivity
and quantum efficiency of the photodiodes. 
The scintillator's wavelength of maximum emission is 425 nm; its light
is shifted by 100 nm towards larger wavelengths (from ``blue'' to
``green'') when re-emitted by the WLS. The photodiode reaches a modest 
sensitivity but high quantum efficiency at this wavelength, 
see Fig. \ref{fig:photodio}.

Finally, the glue between WLS and glass surfaces (n = 1.50) of the 
photodiodes was chosen as an Epoxid resin 
 \footnote{EPO-TEC 302-3M, manufactured by Polytec}
with a refractive index of n = 1.56 which means nearly optimal matching. 
Its transparency in the wavelength range of the WLS is larger than 99 \%.

The attenuation length of a scintillator slab of dimensions 1 x 20 x 200
cm$^3$ is 210 cm \cite{BCsheet1}. This changes drastically for scintillator
strips of dimensions used for the VLQ calorimeter, namely 3 x 5 x 90 (120)
mm$^3$. In these very thin strips the number of reflections which 
scintillation light undergoes, and consequently the net light loss, is
much larger. A measurement of the attenuation \cite{GSchmidt} results in a 
parametrization by a sum of two exponential functions: The region close to 
the exit surface is governed by a very small attenuation length of 2.7 mm, 
the region towards the scintillator centre is described by an attenuation 
length of 101 mm. 

The efficiency of light transmission from scintillators to wavelength
shifters is between 80 \% and 90 \%, the light collection efficiency of
the WLS is about 50 \% \cite{Achim2}. In total, a fraction of 5 - 10 \%
of the light induced by an electromagnetic shower is collected onto the
photodiodes.

\noindent
A survey of the technical parameters of the VLQ calorimeter is given in
Table~1.

\vspace{0.5cm}

\begin{table}[h]\centering
 
\begin{tabular}{|l|c|c}

\hline

         Length                                     & 16.3 cm           \\
         Width                                      & 18.0 cm           \\      
         Height                                     & 15.0 cm           \\
         Volume                                     & 4.4 l             \\
         Mass                                       & 14.2 kg           \\
         No. of tungsten layers                     & 23                \\
         No. of scintillating layers                & 24                \\                
         No. of hor.  scintillating bars/layer      & 18                \\
         No. of vert. scintillating bars/layer      & 24                \\
         No. of hor.  scintillating bars            & 216               \\
         No. of vert. scintillating bars            & 288               \\
         No. of photodiodes                         & 168               \\
         Width of scintillating bars                & 5.0 mm (4.8 mm)   \\
         Thickness of tungsten layers               & 2.5 mm            \\
         Thickness of scintillating layers          & 3.0 mm            \\
         Length of active volume                    & 13.9 cm           \\
         Mean density                               & 8.55 g/cm$^3$     \\
         Depth (X$_0$)                              & 6.7               \\
         Moli\`ere radius (Design)                  & 1.25 cm           \\

\hline      

\end{tabular}                                                                   
\caption{Technical Parameters of the VLQ Calorimeter.}                         
                                            
\end{table}

\subsection{Electronic Read-out}

The read-out via photodiodes requires an electronic
amplification of the diode signals. An energy deposition of 5 GeV 
in the calorimeter liberates roughly 55000 electrons in the photodiodes.
Since these photoelectrons are generally distributed over several
detecting diodes, a charge corresponding to 1000
electrons has to be measured well above noise. 
This condition imposes demanding
requirements on the read-out circuit, especially in terms of electronic
noise. The terminal capacitance of a single diode in the array is
10 pF when operated at a bias voltage of 70 V.

\begin{figure}[hhh]
\begin{center}
\begin{turn}{-90}
\begin{tabular}{cc}
   \includegraphics[width=8cm]{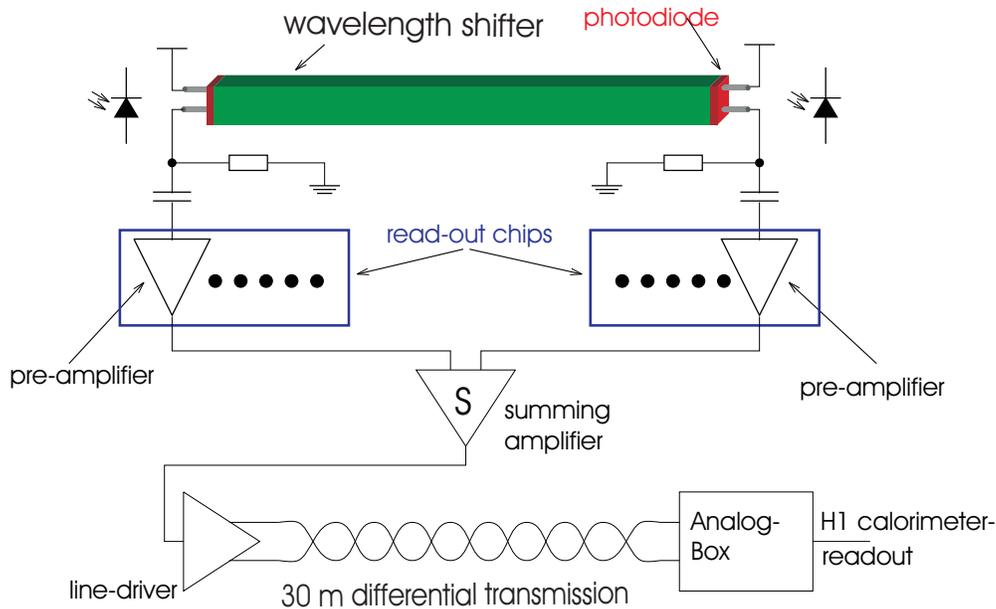}
\end{tabular}
\end{turn}
\parbox{16cm}{\caption{\label{fig:chan} Read-out channel scheme.}}
\end{center}
\end{figure}

\begin{figure}[h]
  \center
     \includegraphics[width=10cm]{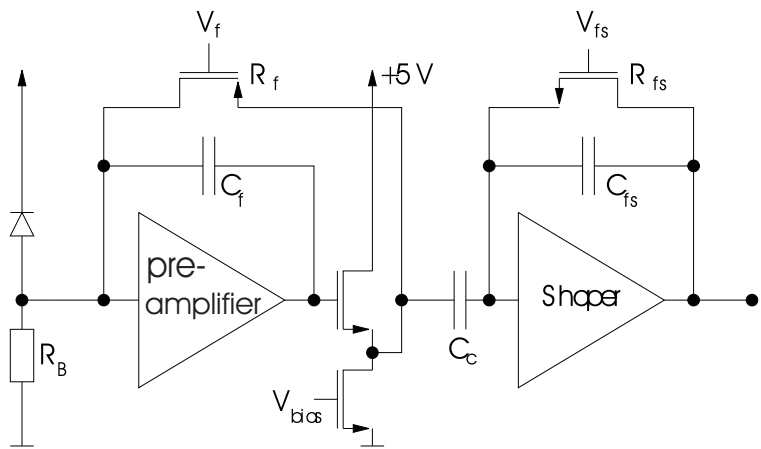}
  \caption{Read-out chip circuit scheme.}
    \label{fig:chip}
\end{figure}

The read-out of the photodiodes is performed via charge sensitive
preamplifiers realized as full custom Application Specific Integrated
Circuits (ASICs), for reasons of compactness and power consumption.
The preamplifier chips are mounted with Chip On Board (COB) Technology, 
on the front and back printed circuit boards 
(PCB) of the calorimeter structure
directly behind the photodiodes, in order to avoid additional capacitive 
loading of the preamplifier inputs through connecting wires. The
amplified diode signals are brought to a top PCB where summing amplifiers
and differential line drivers are located. The preamplifier chip 
was specifically designed and
produced in the Austria Micro Systems (AMS) 1.2 $\mu$m CMOS process.
It features six preamplifier channels with subsequent shaping, fast
trigger summing (see Section 2.5) and an on-chip charge calibration system. 
The rise time of the preamplifier output is 30 ns followed by a 200 ns 
shaping. Principally, the noise goes down with increasing shaping time. 
On the other hand,
the probability of overlapping events increases due to the HERA bunch
crossing distance of only 96 ns. A shaping time of 200 ns was chosen 
as a compromise. The read-out chain and
the scheme of the read-out chip circuit are shown in Fig. \ref{fig:chan}
and Fig. \ref{fig:chip}.

\begin{figure}[h]
  \center
     \includegraphics[width=8cm]{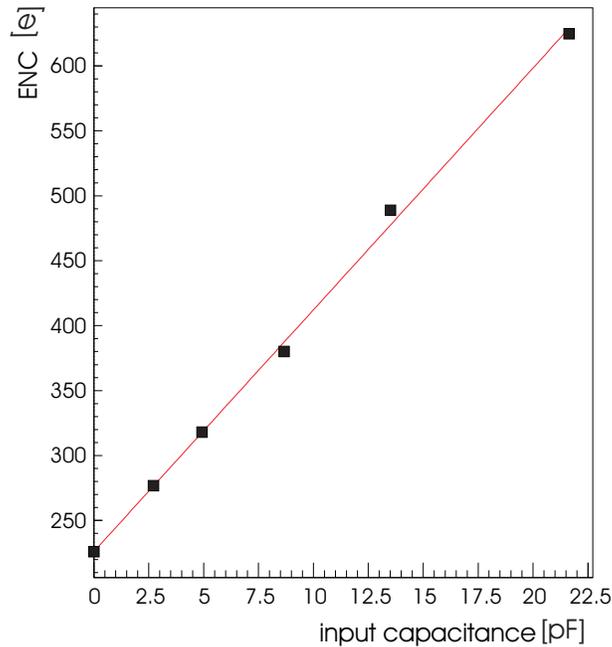}
  \caption{Measured noise (in ENC) as a function of the preamplifier
           input capacitance.}
    \label{fig:noise}
\end{figure}

The noise of
the shaped signal affects energy and spatial resolutions of the
calorimeter. The measured noise of the chip amounts to 226 + 19 x $C$(pF)
electrons r.m.s. where $C$ represents the capacitive load. 
Fig. \ref{fig:noise} shows the noise measured at the output of the charge
amplifier of the read-out chip as a function of the capacitance at
the input. 

\noindent
With the
photodiodes used here, this corresponds to an effective noise figure
of 416 electrons r.m.s. With an estimated signal of 11 000 electrons for 
a 1 GeV electron shower out of which 75 \% are deposited in a core seen
by two horizontal and two vertical scintillator bars with front and back
read-out, a calibrated noise figure of 120 MeV is anticipated.

After shaping, the two signals from either end of one WLS bar are summed
and sent, through a 30 m long cable, differentially to the
analogue receivers located in the electronics hut of the H1 experiment
This system is identical to that one operated in the read-out of
the H1 backward calorimeter BEMC \cite{BEMC}.
The read-out and amplification electronics, including the
line driver, are located inside the VLQ calorimeter housing. 
The total power consumption of 25 W is mainly due to the line drivers, and 
requires water cooling of the read-out electronics.

The calorimeter
signals are re-transformed into unipolar signals by a differential
receiver. The signal is delayed by 2.2 $\mu$s with the help 
of an adjustable analogue delay line.
This time is needed to store the signals until the first level trigger of the
H1 detector has made a decision. The signals from this analog pipeline
are amplified once more and, in case of a positive trigger decision, stored
in a Sample-and-Hold circuit as analog voltages. The Sample-and-Hold
circuit stores the voltage at exactly that time when the trigger signal
arrives. This requires a precise timing of the calorimeter signals 
whose maxima are to be stored upon trigger signal arrival.

One analog receiving card accomodates 16 equally built channels. The voltages
stored in these 16 Sample-and-Hold circuits are, after trigger decision,
transferred via a multiplexer to the H1 standard calorimeter read-out
system \cite{H1Calo} where signals are digitized in a 12 bit 
ADC~\footnote{Analogue-to Digital Converter}, and 
stored. In this way, the 84 signal and 4 trigger channels of a 
calorimeter module are read out. The trigger channels are also fed into
the analog read-out in order to perform trigger control measurements.

\subsection{The Calorimeter Trigger}

The VLQ spectrometer is devoted to new physics regimes where high energetic
particles are scattered through small angles into the backward region 
of H1, i.e. the region downstream the electron beam. The VLQ becomes 
therefore an important and independent element of the H1 trigger system.

A VLQ calorimeter trigger signal can carry information on energy deposited,
above a certain threshold, in a certain section of the calorimeter.
A maximum number of four trigger signals per module is possible as dictated 
by the resources of the H1 Central Trigger Logics (CTL).
The signal has to be derived from the (2 x 18) + (2 x 24) read-out channels
available. The most simple conceivable trigger signal would be just the sum 
of the signals of all read-out channels. This straight - forward
solution has two drawbacks: First, since an electromagnetic shower 
distributes its energy over few channels only, signals are added from 
calorimeter regions ``far away'' from the shower, thus adding just noise
to the sum signal. Second, a rather frequently occurring background effect 
known as ``Single Diode Effect'' (or ``Nuclear Counter Effect'') would not
be suppressed in such a simple scheme.

The calorimeter noise can be separated into two classes: Coherent and
incoherent noise. The former is due to signals, e.g. out
of the electronic environment, which are picked up by all calorimeter
channels simultaneously. They will lower or raise the original signals
by roughly the same amount. Incoherent noise is individual (e.g. 
statistically generated) activity, with a Gaussian amplitude spectrum 
centered at zero whose spread is supposed to be reduced when summed over. 
The contribution from coherent noise is proportional to the number of
noisy channels, the contribution from incoherent noise to its square
root. To minimize the noise contribution, in particular from coherent
noise, the number of channels 
to be summed up to yield a signal must be kept as small as possible.

The ``Single Diode'' (SD) effect is generated by particles passing 
through the 
depletion layer of a diode and leaving an amount of energy which mimics 
a high energy shower signal. 

The suppression of
both coherent noise and SD effect require a more segmented trigger scheme
because the ``total sum'' solution would lead to an intolerable trigger
threshold behaviour, and suppression of SD events needs comparison of
channels opposite to each other. 
Therefore, in a first step only calorimeter signals
from horizontal scintillator bars, read out at the vertical sides, are
used for triggering. They suffer less strongly from the SD effect because
the calorimeter regions close to the beam line, and thus the corresponding
horizontal photodiode arrays, are much more frequently hit by scattered
electrons due to their $1/Q^4$ (i.e. Rutherford) angular distribution. 

\begin{figure}[h]
  \center
     \includegraphics[width=8cm]{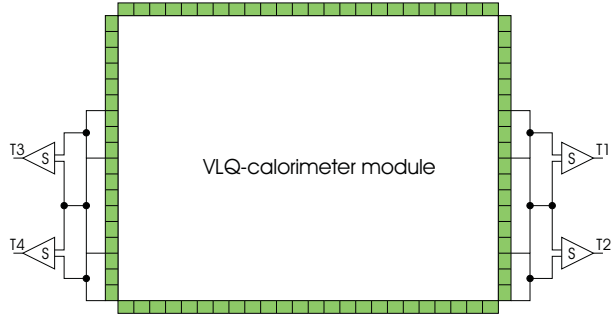}
  \caption{The VLQ calorimeter geometrical trigger pattern.}
    \label{fig:trigpat}
\end{figure}

In a second step, the calorimeter module is logically divided into trigger
segments. Vertically,
two windows are constructed of nine channels each, with an overlap of six
channels. The overlap region is chosen large enough to cover a complete
shower which deposits nearly all its energy in one central plus two left
and two right neighbouring channels. The trigger efficiency is thus not
reduced for particles impinging close to the edge of a window. The trigger
pattern is shown in Fig. \ref{fig:trigpat}.

Each vertical window is read out at
either side so that ``left signal'' and ``right signal'' can be 
compared. As the light attenuation in the scintillator strips is weak,
a real shower should lead to roughly equal signals at either side, while
SD events do not, and thus can easily be recognized at the trigger level.

The outmost six channels (which point towards the main detector) are 
not included in the trigger because they are hidden in the shadow of 
the SpaCal insert \cite{SpaCal} which was supposed to be modified in 
a later upgrade step. The trigger scheme is, as is the read-out 
scheme, symmetrical for both modules. In summary, four analog trigger 
signals per module, or eight signals for the entire spectrometer, are 
available.

The analog trigger signals are delivered to the trigger module which
prepares the digital trigger information. First, the analog signals from
opposite sides of a window are added in order to avoid inhomogeneities in
the trigger efficiency. In addition, thresholds are applied to the eight
original sums yielding the signals N1 to N8; these (low) thresholds are
labelled ``noise triggers''. Furthermore, two independent (``low'' and
``high'') thresholds are applied to the newly built sums, yielding
another eight digital signals (L1+3 to L6+8, H1+3 to H6+8). These digital 
signals 
are synchronized with the HERA clock on the trigger module. The 16
digital signals generated on the trigger module are fed into a 
general purpose digital input - output card.
It accomodates a freely programmable look-up table which
combines the input signals to the eight VLQ trigger elements delivered to
the Central Trigger Logics of H1. All thresholds in the trigger
module can be adjusted individually. 

Finally, there are eight VLQ - derived trigger elements, four derived
from the top module, and four derived from the bottom module. For
symmetry reasons, only the four top - derived elements are explained
here:
\\
\newline
  ``VLQ-top-noise'' = (N1 AND N3) OR (N2 AND N4)  
\newline
  ``VLQ-top-low'' = (L1+3) OR (L2+4)  
\newline
  ``VLQ-top-high'' = (H1+3) OR (H2+4)
\newline
  ``VLQ-top-SDE'', a trigger element which signals the presence of
  asymmetric energy deposition, i.e. a so called ``single diode'' event.
\\  
\newline
\noindent  
The corresponding thresholds (``noise'', ``low'', ``high'') are roughly
3.5 GeV, 7.0 GeV, and 10.0 GeV, respectively. The VLQ - derived trigger
elements are combined with each other, and trigger elements from other
H1 subdetectors, in the Central Trigger Logic.

\section{The Cluster Reconstruction Algorithm}

In order to reconstruct physical quantities from 
calorimeter data, an energy correction and cluster
finding algorithm to evaluate the raw data has been developed. The
calorimeter raw data consists of the digitized signal amplitudes
measured in each of the 168 calorimeter channels and stored in units 
of ADC counts. In addition
the amplitudes of the eight trigger summing channels are recorded.

The reconstruction is subdivided into two separate software
modules which are executed sequentially. The first module performs
operations upon individiual channels, such as the transformation
of ADC counts into energies in GeV, while the second
module locates compound objects, the ``clusters'', which are
constructed from logically connected channels. Since the VLQ
calorimeter is part of the H1 four-level triggering system, the
reconstruction is developed such that the algorithms can be
executed at the fourth trigger level, a multi-processor
computing farm. This restricts the complexity of the
algorithms to be run and the size of the reconstruction software because
only a limited amount of time to evaluate the event is available,
and constraints upon the hardware resources exist. 

\begin{figure}[h]
  \center
     \includegraphics[width=14cm]{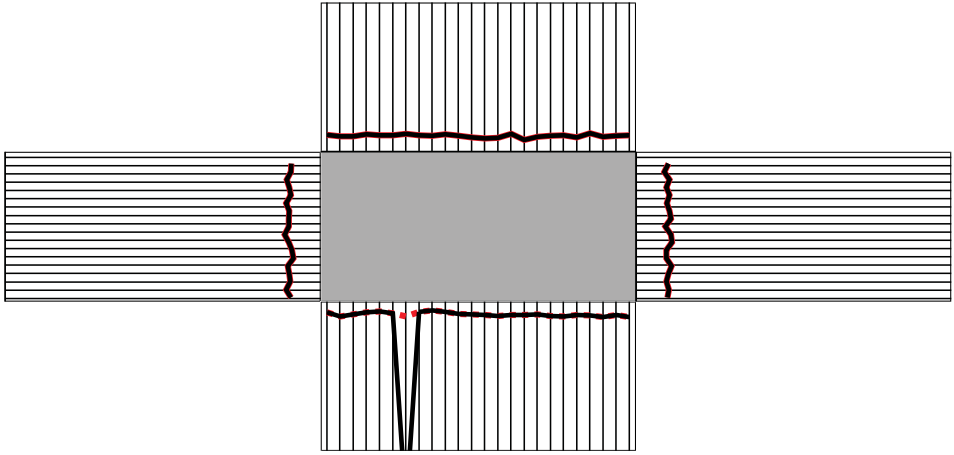}
  \caption{Projected energy deposits of a ``single diode'' (SD) event.}
    \label{fig:SDE}
\end{figure}

The first step in the
reconstruction process is unpacking of the bitwise packed
calorimeter raw data. One raw data word contains the channel
energy and the corresponding channel identifier. After unpacking,
the pedestal for each channel, obtained as the average from an
offline random trigger data set and stored in an online-accessible
data base, is subtracted from the read out signal. 
The individual channel-to-channel alignment constants, 
also determined in an offline
calibration as described in Section 5, are applied. 
Thereafter the
second module, the cluster construction, is executed. In order to
be able to reconstruct clusters, potential ``single
diode'' channels (SDs, see Section 2.5) have to be removed. 
A display with the four projections of a typical single diode event 
is shown in Fig. \ref{fig:SDE}.

\noindent 
In view of the projective read out of the
VLQ calorimeter, the removal of SD channels is a straight-forward 
task. Both the horizontal and the vertical scintillator
strips are read out at either end. Thus there are two independent
energy measurements of each projection, and one can compare
the two signal amplitudes from the two ends of a scintillator bar. 
If the energy in one channel is
significantly larger than that in the corresponding opposite one, 
the channel is treated as a single diode one, and its
energy is set to zero.

However, the situation is generally a bit more complicated
because a ``single diode'' and a ``real'', i.e. scintillator light induced, 
event might overlap. Therefore the two neighbouring channels of
the suspected SD channel are checked if their amplitudes
exceed a certain level above noise. If this is not the case, the
central channel is treated as an SD channel and its energy is reset to zero. 
If the amplitudes are well above noise, they are averaged, and the
mean value is assigned to the central channel. This procedure cannot
be applied, however, if the single diode candidate is close to one
of the calorimeter edges. In this case its amplitude is replaced
with that of its opposite partner. SD channels are detected in roughly
25 to 30~\% of all events, depending on the running conditions.

After the check for and removal of SDCs, a two step
cluster construction is performed. The first step aims at finding
``pre-clusters'' independently in each of the four projections. To
recognize a pre-cluster, the prescriptions of simple
calculus to find minima and maxima of an analytic function are 
applied and transposed in order to handle discrete values as those
of calorimeter channel amplitudes. The first and second derivatives
are approximated as
\begin{eqnarray}
A'(i) &=& \frac{A(i+1)-A(i)}{h}    \\
%A''(i) &=& \frac{A(i-1) - 2 \cdot A(i) + A(i+1)}{h^2}.
A''(i) &=& \frac{A(i) - 2 \cdot A(i+1) + A(i+2)}{h^2}.
\end{eqnarray}
Here, $h = (i +1) - i = 1$ represents the distance between adjacent
channels. Before calculating the derivatives, a three $\sigma$ noise cut is
applied on the individual channels. This reduces  the building of
noise clusters due to random fluctuations in the amplitude distribution.

\noindent 
After calculating the derivatives, the channels with minima
and maxima in the amplitude distribution are flagged. 
This is done for each of the four projections independently without using 
any knowledge from the other projections. The projective readout is
then used to cross-check the results by using information from the
opposite projection. The numbers of minima and maxima found are
compared for each corresponding projection. In the case the numbers 
do not match, the region in the projection where an extremum was 
expected, but none was found, is re-examined. This case is quite
frequent and is not due to a malfunction of the algorithm, but rather
due to scintillator attenuation and cut-off effects. In this step, the
noise cut is no longer applied, and the pre-cluster is verified if the
channels around the maximum contain at least 40 \% of the maximum
amplitude. 
In this case a pre-cluster will be constructed in the projection
where previously none was found. To achieve consistency in the
number of pre-clusters found in corresponding projections is 
important because in the final clustering procedure the pre-clusters
are combined to form final clusters.
In order to verify a cluster, all four projections must match and are
needed to reconstruct both x- and y- coordinates.

The situation is somewhat more complicated if two or more particles
enter the calorimeter volume. In the case of two clusters (the rare
case of three or more clusters can be neglected), there are two
possible configurations: The clusters are aligned in either $x-$ or
$y-$ direction, or they are distinct. In both cases a pre-cluster
construction algorithm is applied analogous to the one - cluster case,
with a few assumptions in order to define a cluster separation 
criterion \cite{Nix}, so that channel energies can be properly 
assigned to clusters.

As soon as pre-clusters are identified and their energies determined,
the energy-weighted coordinates of pre-cluster cells in all four
projections are used to calculate the coordinates of the cluster 
centre of gravity. Following experience with the large backward 
calorimeter SpaCal \cite{SpaCal}, a logarithmic weighting procedure
is applied. The individual weights of contributing channels $i$ are 
determined as
\begin{equation}
    w_i = \max(0,W_0 + \ln(E_i / \sum_i E_i)) ,
\end{equation}
so that the coordinates of the cluster centre of gravity become
\begin{equation}
   x_{log} = \frac{\sum_i x_i w_i}{\sum_i w_i} .
\end{equation}
$W_0$ is a dimensionless cut-off parameter based on assumptions 
on the the shower profile \cite{Achim2}. Introducing $W_0$
has the effect that only channels above a certain threshold are
included in the calculation of the shower centre of gravity. 
Finally, the cluster radius is calculated as
\begin{equation}
  R_{cluster} = \frac{\sum_{i}^{N} E_i \cdot \sqrt{(x_{cl} - x_i)^2}}{E_{cl}}
\end{equation}
where N is the number of channels associated with the
clusters, $x_{cl}$ the cluster center of gravity, and $E_{cl}$ the
cluster energy.

After reconstructing the pre-clusters in the four projections, the
final (full) cluster properties are determined by adding the pre-cluster
energies and averaging the coordinates of the pre-cluster centres.

In case of more than one cluster per projection the situation is more 
difficult, as illustrated in figure \ref{fig:twocl}:

\begin{figure}[h]
  \center
      \includegraphics[width=14cm]{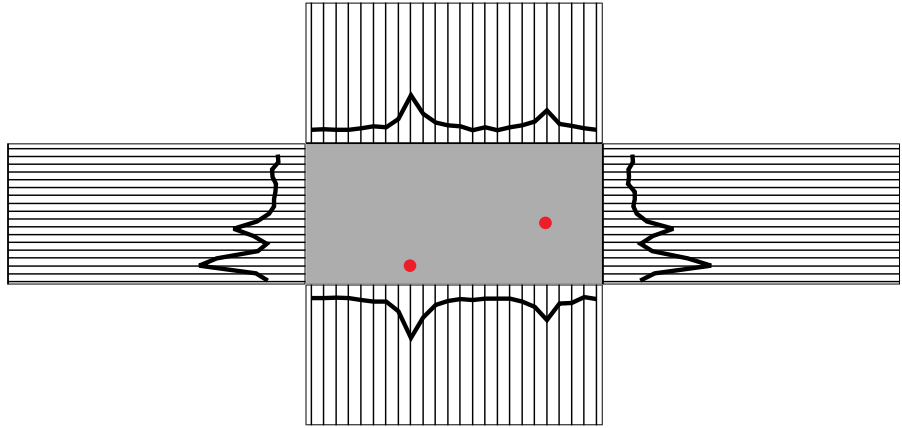}
  \caption{Projected views of a two-cluster event.}
    \label{fig:twocl}
\end{figure}

Two particles hitting the calorimeter lead to eight pre-clusters from
which four clusters can be reconstructed so that there is no unique 
solution. If however the two incident particles deposit clearly different
amounts of energy in the calorimeter, combining the low energy pre-clusters
versus the high energy ones allows a non-ambiguous solution.
This method will of course fail if the particles have similar energies. 
In this case no unique solution is possible, and the decision needs
to be left over to the VLQ track detector \cite{Hurling}.

\section{The Calibration Concept}

\subsection{Test Beam Calibration}

Both VLQ calorimeter modules were tested in an electron 
test beam at the Deutsches Elektronen - Synchrotron DESY in Hamburg. 
The beam, supplied by the DESY synchrotron, had energies adjustable
between 1 and 6 GeV. 

In the test setup, the VLQ module to be investigated was fixed at the
moving mechanism constructed for operation in H1. As this device supplies
only vertical movement, horizontal displacements were performed on a
support table driven by a step motor. The vertical and horizontal
displacement steps were 5 mm each, corresponding to the scintillator
strip width.
 
The distance 
between beam collimator and calorimeter module was 5.70~m. The
position of the beam in a vertical plane in front of the calorimeter
was defined with the help of a silicon tracker telescope. The telescope
consisted of two densely packed bundles, each equipped with four Si strip 
detector planes. The strip orientations were y - x - y - x - x - y - x - y.
Each plane contained 384 Si strips of 50 $\mu$m width. The distance between 
the layers inside one bundle was 10 mm, the two bundles were 17 cm apart from
each other. The cross section covered by the telescope was 2 x 2 cm$^2$. 
The Si - telecope was read out sequentially by a 6 - bit CAMAC ADC.
It was sandwiched between two scintillation counters which, in coincidence
with the telescope signals, provided a trigger pulse. The read-out chain
for the VLQ calorimeter modules was identical to that one used in H1 with 
the exception that the synchronization between test beam generated events
and the HERA clock which controlls the readout had to be imposed
artificially \cite{Achim2}.

\begin{figure}[h]
 \begin{center}
   \includegraphics[width=13cm]{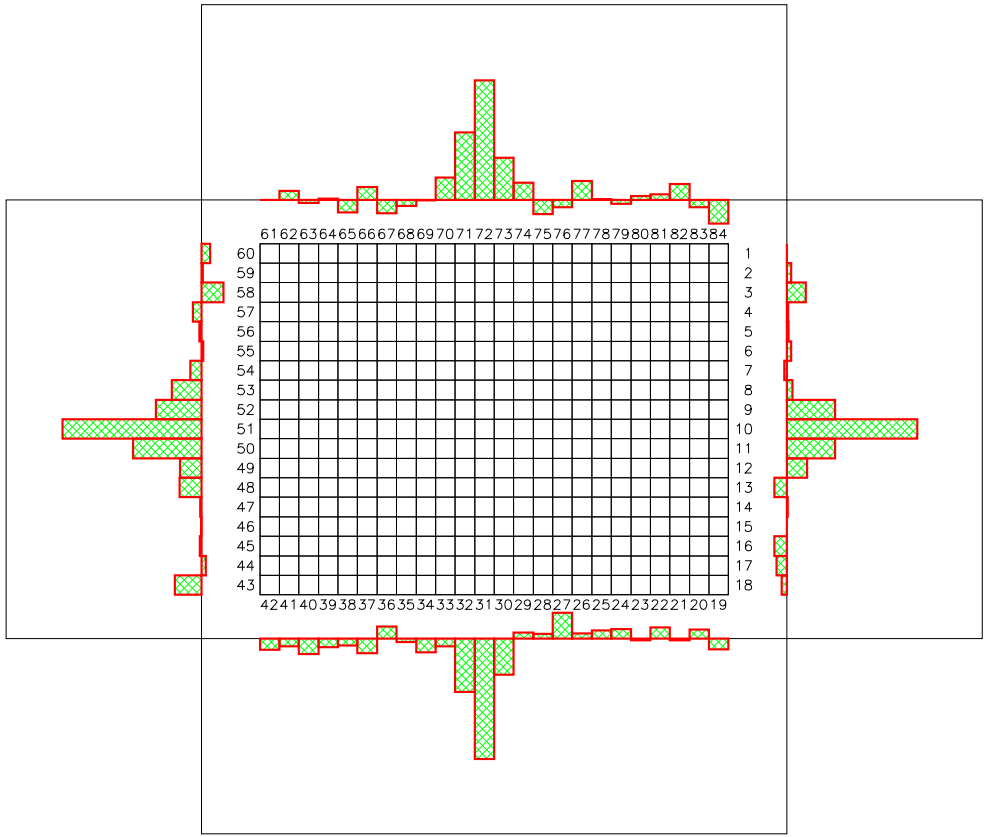}
  \begin{quote}
  \caption{Energy deposit of a 6 GeV electron in the VLQ calorimeter,
           recorded in a test beam.}
    \label{fig:testev}          
  \end{quote}
 \end{center}
\end{figure}

Fig. \ref{fig:testev} shows a test beam event as a fourfold 
projection of the energy
deposition in the calorimeter. Each of the four histograms displays the
(pedestal corrected) contents of the corresponding ADC channels. To
determine the shower energy, the entries of the ADC channels of all four
histograms have to be summed up.

In an ideal homogeneous sampling calorimeter of infinite depth, the output 
signal is proportional to the incident and completely absorbed energy. The 
proportionality factor which converts ADC counts into energy is called 
calibration constant. 

A real calorimeter is neither perfectly homogeneous nor infinitely deep.
The response will depend on the impact position of the primary
particle which, together with unavoidable leakage effects, deteriorates the
energy (and spatial) resolution. Calibration has therefore a twofold
purpose: In a first step, the response of different channels to identical
energies must be equalized (``intercalibration''), and in the second 
step the conversion factor to obtain absolute energies must be determined. 

\begin{figure}[h]
  \center
     \includegraphics[width=8cm]{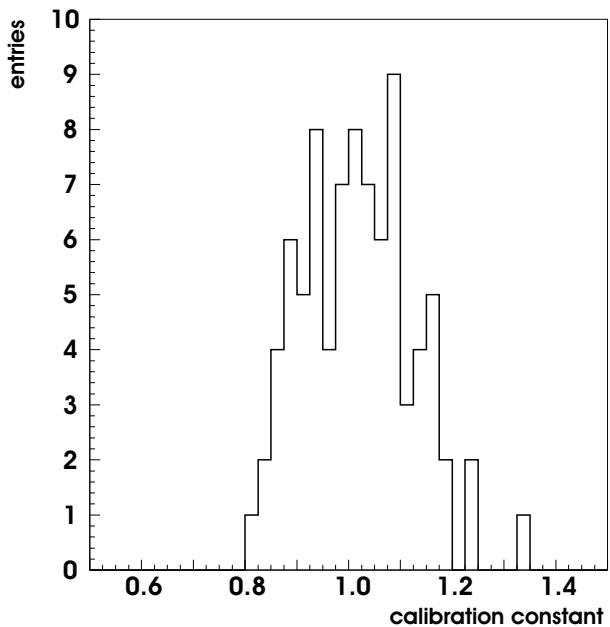}
  \caption{Distribution of calibration factors from test beam measurements.}
    \label{fig:calconst}
\end{figure}

Intercalibration is performed through two independent 
spatial scans, one along the
horizontal side of the calorimeter module, and one along the vertical one.
Channel-to-channel responses vary due to inhomogeneities which are mainly
caused by imperfect light connections. After a first go-through, the 
individual single channel responses can be pre-equalized by appropriate
weights. In the subsequent step, the spread of output signals will be
reduced but still be present, mainly due to light cross talk at the WLS -
photodiode connections. Therefore the procedure is repeated until a
convergence criterion is fulfilled. In the end, after application of the
overall calibration constant to convert Volts into GeV, the individual
calibration factors determined for one module have a mean close to one 
and an r.m.s. of about 10 \%, see Fig. \ref{fig:calconst}.

\begin{figure}[htbp]
  \begin{center}
    \centerline{
      \includegraphics[width=7.5cm, height=6cm]{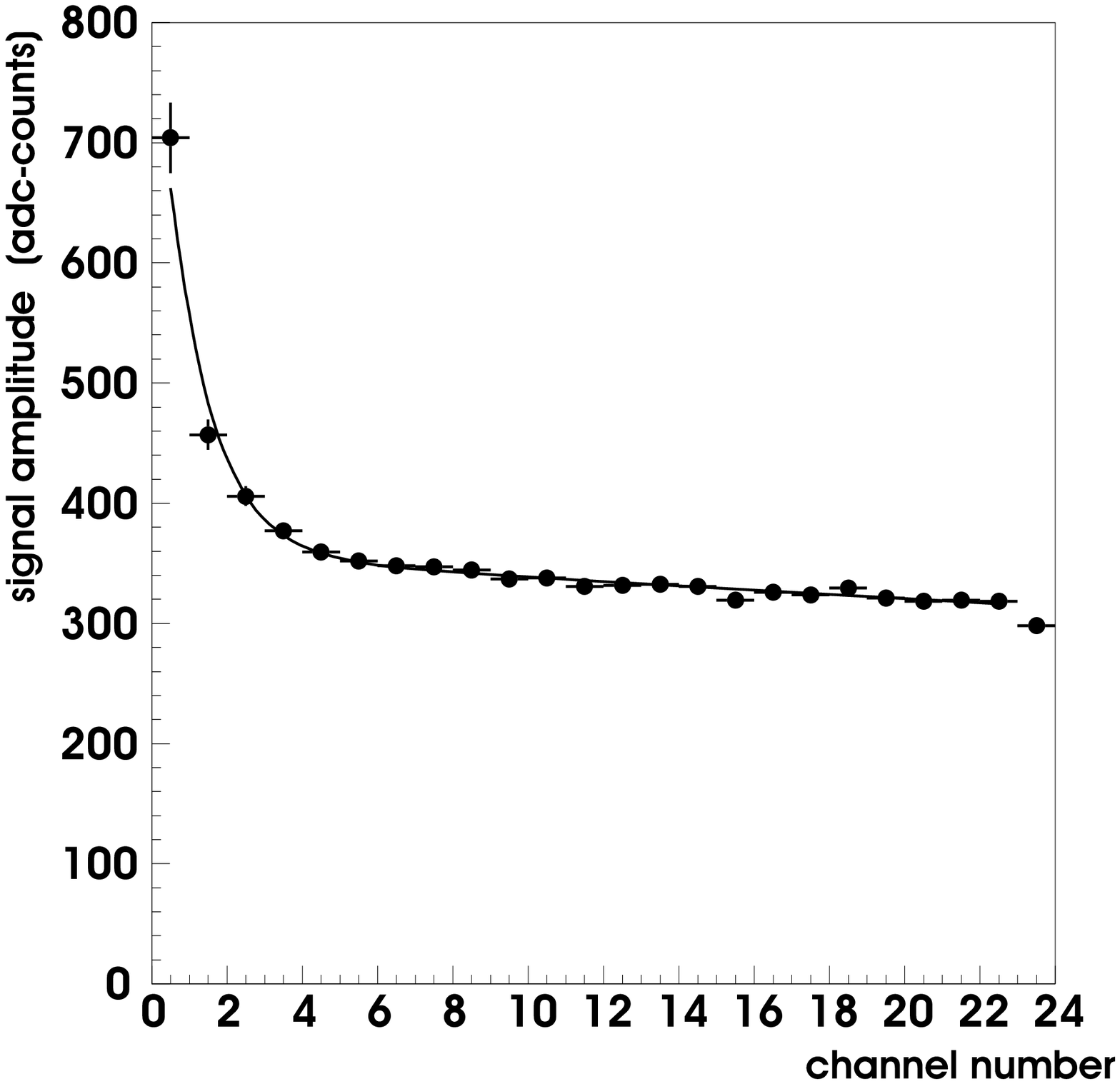}
      %\quad
      \includegraphics[width=7.5cm, height=6cm]{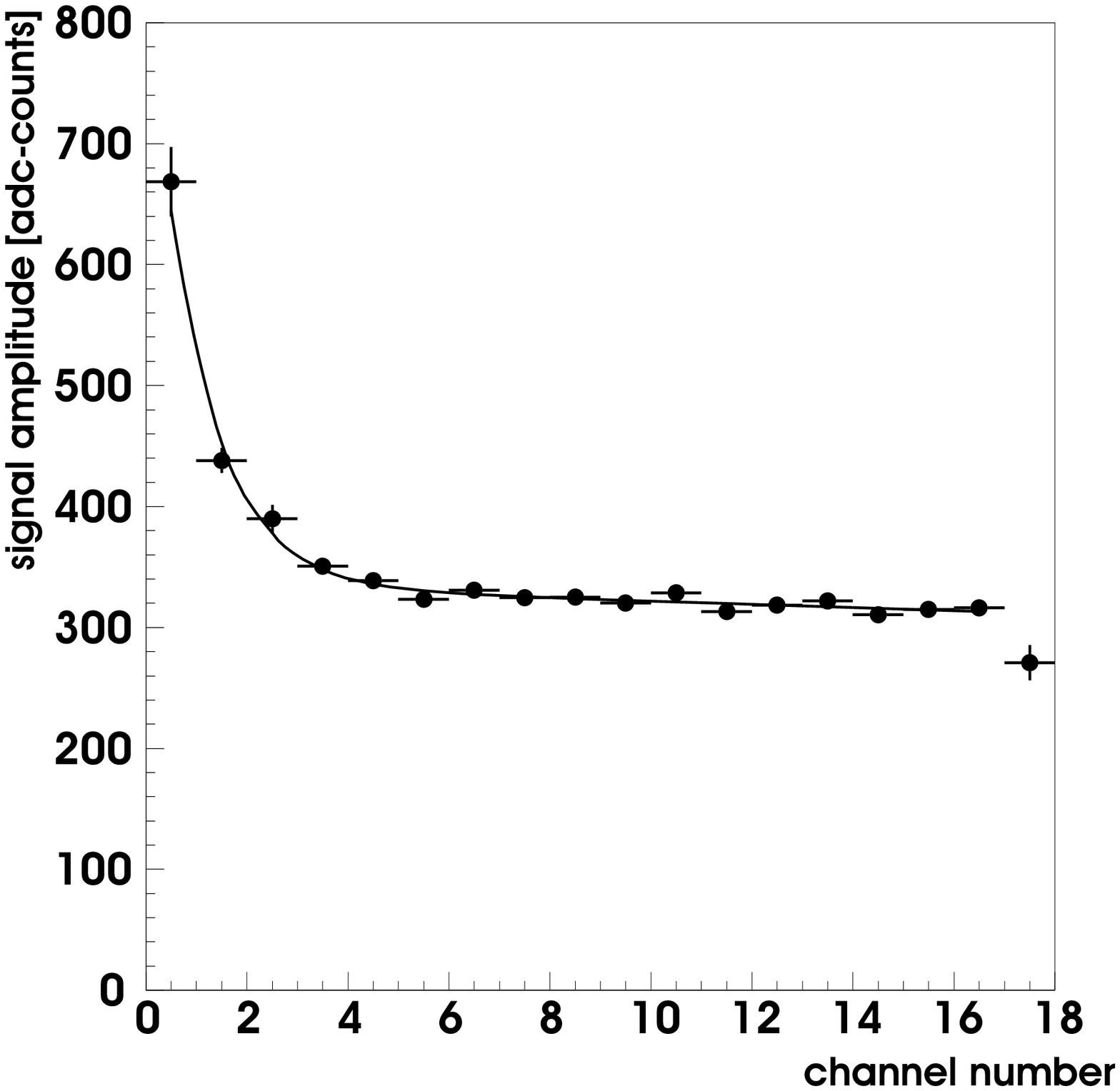}
      }
    \begin{quote}
      \caption{Light attenuation as a function of the scintillator
                strip hit: a) vertical, b) horizontal strips.}
      \label{fig:attenu}
    \end{quote}
  \end{center}
\end{figure}

As mentioned above, the scintillator strips of the VLQ calorimeter show
an absorption length much smaller than the corresponding bulk material.
This affects the energy reconstruction and requires correction. The light
attenuation process was quantified through a test series where the beam
impact was varied along a horizontal (vertical) layer of strips, and the
corresponding vertical (horizontal) WLS array was read out. The results
are shown in Fig. \ref{fig:attenu} a and b, and where the ADC signal height 
is plotted as a function of the number of the vertical (a) and horizontal (b) 
strip hit.

Apparently, the extinction curve shows two
components and can be fit, up to the penultimate point, by the sum of 
two exponentials:
$$ F(d) = C_1~\exp(-\lambda_1~d) + C_2~\exp(-\lambda_2~d), $$
where $d$ denotes the horizontal or vertical distance between beam impact
point and light connection, respectively. The reduced light yield from 
the last strip hit is due to shower particle losses at the calorimeter 
edge. The 
enhanced light yield close to the connection scintillator - WLS is 
caused by its geometry which allows for a wider emission angle at this 
junction. For the extinction parameters, the fit delivers the results
$\lambda_{1,hor} = (0.0128 \pm 0.0008)$~cm$^{-1}$, 
$\lambda_{2,hor} = (1.8 \pm 0.2)$~cm$^{-1}$, 
$\lambda_{1,vert} = (0.012 \pm 0.002)$~cm$^{-1}$, and
$\lambda_{2,vert} = (2.0 \pm 0.2)$~cm$^{-1}$.

\noindent
The light extinction is corrected for when applying the final
calibration constants.

\begin{figure}[h]
 \begin{center}
     \includegraphics[width=8cm]{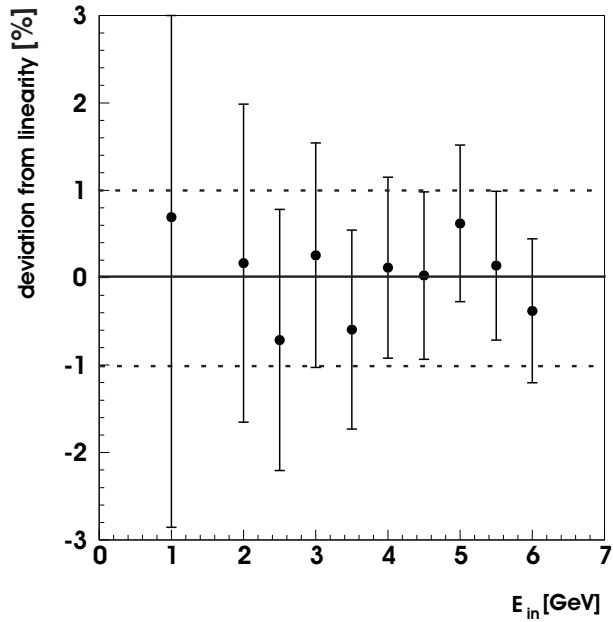}
  \begin{quote}
  \caption{Relative deviations from calorimeter linearity, from test beam
           measurements in the energy range from 1 to 6 GeV.}
    \label{fig:devialin}
   \end{quote}
 \end{center}
\end{figure}

After an energy scan, the linearity behaviour of the calorimeter can be
studied in the available energy range from 1 to 6 GeV. The result is
displayed in Fig. \ref{fig:devialin} which shows the relative deviation from
linearity. Deviations are below 1 \% and consistent with the simulation
results (cf. Fig. \ref{fig:lineasim}).

An additional result of the energy scan is the energy resolution, which,
parametrized as in section 2.1, reads
\begin{equation}            
  \frac{\sigma(E)}{E} = \sqrt{\left(\frac{(19 \pm 6) \%}
                             {\sqrt{E/\text{GeV}}}\right)^2
                  \oplus ((6.5 \pm 3) \%)^2
                 \oplus \left(\frac{(23.4 \pm 0.9)~\%}{E/\text{GeV}}\right)^2}.
\end{equation} 
This is marginally consistent with the simulation result of eq. 2; 
differences are due to the limited (low) energy range covered by the 
test beam and its impurities, e.g. due to secondary scattering processes,
which tend to distort the energy measurements 
\cite{Achim2}.

The reconstruction of cluster spatial coordinates is also based on the energy
measurement: The shower centre of gravity in either projection is
expressed as energy weighted mean of scintillator strip coordinates which
by construction have a digitization of 5 mm. For consistency, the same
five channels are used for spatial reconstruction as for shower energy 
reconstruction. 

\begin{figure}[h]
  \center
     \includegraphics[width=8cm]{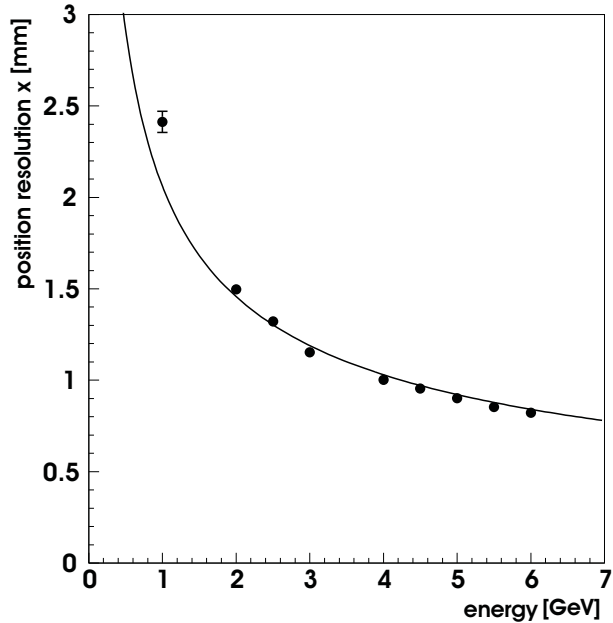}
  \caption{Calorimeter spatial resolution, x - coordinate.}
    \label{fig:xresol}
\end{figure}

The spatial resolution of the calorimeter is determined by comparing the
reconstructed beam impact position (according to the procedure described 
above) and the impact position defined with the help of the silicon tracker 
matrix whose resolution is roughly 0.1 mm. The result for a beam energy 
of 4 GeV, for either calorimeter projection, is about 1 mm. The resolution
for the horizontal (x -) coordinate is shown in Fig. \ref{fig:xresol}. 
When parametrizing the spatial resolution
as a function of energy, a fit of the form
$$  \sigma(x) = \frac{P_1[\text{mm}]}{\sqrt{E/\text{GeV}}}, $$
yields $P_1 = (2.06 \pm 0.02)$ mm. This implies a resolution of better than
one millimeter for energies above 4 GeV.

\begin{figure}[h]
\begin{center}
     \includegraphics[width=8cm]{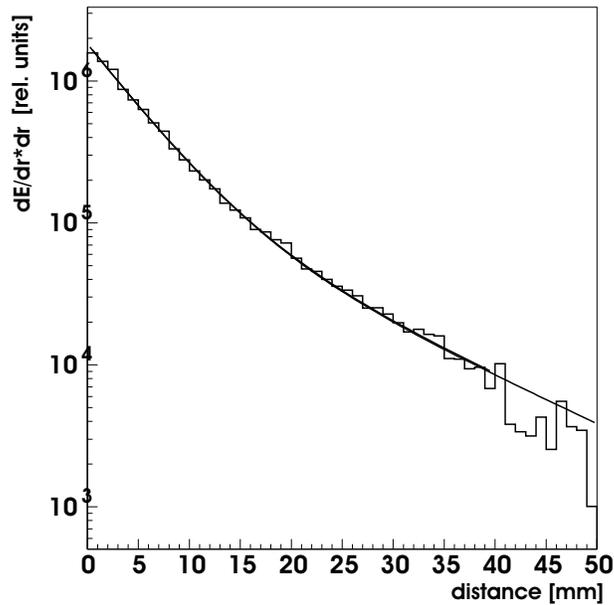}
 \begin{quote}
  \caption{A transverse shower profile, i.e. energy deposition as a 
           function of distance from the shower axis, at an energy of 
           4 GeV as determined in the test beam.}
    \label{fig:shopro}
  \end{quote}
 \end{center}
\end{figure}

The projective read-out structure of the calorimeter finally allows for a
direct sampling of the lateral shower profile. This is performed by
measuring a large number of showers in the central area of a calorimeter
module, at an energy of 4 GeV. In a large number of measurements,
noise and statistical 
fluctuations in the shower profile cancel. The bin size is given by the 
resolution (about 1 mm). In Fig. \ref{fig:shopro} all 
shower profiles are plotted 
versus the radial distance from the shower axis such that each shower 
centre is shifted to R = 0.
One recognizes a two - component structure of the profile which
can be parametrized as the sum of two exponentials. A fit to the 
distribution of Fig. \ref{fig:shopro} yields, for the two parameters 
for the ``fast'' and ``slow'' lateral damping, the values 
$(2.2 \pm 0.1)$ cm$^{-1}$ and $(0.77 \pm 0.05)$ cm$^{-1}$.

\noindent
Integrating the exponential profile function with these parameters
yields a Moli\`ere radius of 1.5 cm which is slightly larger than the
value predicted by the simulation (1.25 cm, see section 2.1). The 
difference is explained by optical cross talk between neighbouring 
scintillator 
and read-out channels which leads to an apparent shower broadening,
and which was not taken into account in the simulation.

The following Table 2 gives an overview over the most important parameters
quantifying the VLQ calorimeter's performance in the test beams.

\vspace{0.5cm}

\begin{table}[h]\centering
 
\begin{tabular}{|l|c|c}                 
\hline

Energy resolution, sampling term         & (19  $\pm$ 6) \%          \\
Energy resolution, constant term         & (6.5 $\pm$ 3.0) \%        \\
Energy resolution, noise term            & (23  $\pm$ 1) \%          \\
Deviation from linearity                 & $<$ 1 \%                  \\
Horizontal (vertical) spatial resolution & $<$ 1 mm for E $>$ 4 GeV  \\
Moli\`ere Radius                         & 1.5 cm                    \\     
                                                                                                                                                          
\hline      

\end{tabular}                                                                   
\caption{Measured VLQ calorimeter performance parameters.}                         
                                            
\end{table}

\subsection{Normal Calibration Using the ``Kinematical Peak''}

In order to calibrate the VLQ calorimeter during the standard 
experimental data taking
phase and at high energies, the observation of the ``kinematical peak''
in electron - proton scattering 
is exploited: In the energy spectrum of electrons scattered off protons
with moderate $Q^2$, a peak appears at the position of the incident 
electron beam energy. The response of a calorimeter can be adjusted
accordingly.

The calibration proceeds in several steps: After selecting the kinematical
peak events, a channel-to-channel intercalibration is performed in order
to homogenize channel responses. This is followed by adjusting the
absolute energy scale in terms of position dependent calibration factors.

\subsubsection{Event Selection and Channel-to-Channel Intercalibration}

\begin{figure}[h]
  \center
      \includegraphics[width=14cm]{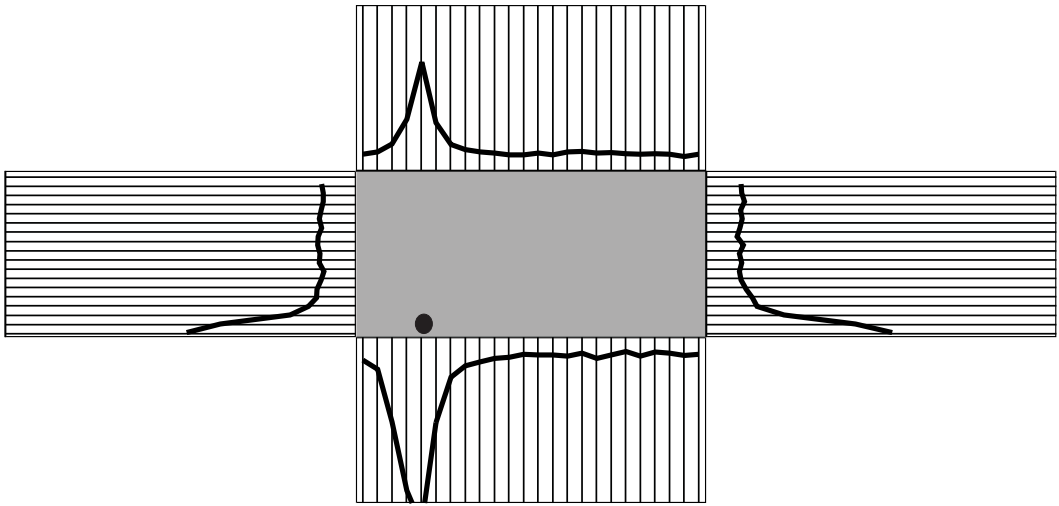}
  \caption{Projected views of a ``Kinematical Peak'' event.}
    \label{fig:kinPeakEvent}
\end{figure}

Events to be used for the kinematical peak calibration must fulfil
the following selection criteria: The electron must deposit an energy above
the highest trigger threshold in one of the calorimeter modules;
the value of the ``inelasticity'' $y$ reconstructed 
following the prescription 
by Jacquet and Blondel \cite{JB} which indicates hadronic energy 
deposition in the central calorimeter must obey $y_{JB} \leq 0.04$;
as at HERA the electron beam energy is $27.56\,\text{GeV}$, only clusters 
with energies $E_{cl} \geq 20.0\,\text{GeV}$ are accepted. A requirement
of the reconstruction software is that exactly one cluster and no SD
channel was found, or at least no SD channel which is
connected to the cluster.

The aim of the channel-to-channel intercalibration is to ensure
that equal energy depositions result in equal responses of
all channels. Variations in the channel energy responses can 
have different sources, the most important ones are due to
optical and electronical cross talk. 
The optical cross talk is due to light emission into neighbouring
photodiodes, caused by overlapping light cones in the thin glass 
pane between wavelength shifters and photodiode surfaces.

The channel-to-channel calibration is a two step procedure
whose principle has been sketched in Section 4.1 . The algorithm
used for the normal calibration is briefly described below:

In a first step the signal of the channel with the
maximum energy deposition is filled into a histogram associated
to that specific channel, and the mean for each channel is 
determined. In order to calculate calibration factors, a
reference value is needed which in this case is the ``mean of
(the above defined) means'', averaged over all channels belonging
to one wavelength shifter array. This procedure is expressed in 
the following equation:
\begin{equation}
M^{(1)} = \frac{ \sum_{i,N_{i}\geq N_{limit}} m_i^{(1)}}{n^{(1)}.}
\end{equation}
Here, $n^{(1)}$ is the number of channels entering the average after
the first iteration, and $m_i^{(1)}$ is the mean value of the
histogram which contains the measured energies of channel $i$.

$M^{(1)}$ is the global mean in an given wavelength shifter array
for the first iteration. The effect of the calibration is
such that after applying the first iteration calibration
constants 
\begin{equation}
f_i^{(1)} = \frac{M^{(1)}}{m_i^{(1)}.}
\end{equation}
the mean energy in any channel is shifted to the global mean.
This procedure is then repeated until the change in the mean
energy of any channel with respect to the previous iteration
is smaller than $0.5 $\%.

\begin{figure}[h]
  \center
     \includegraphics[width=14cm]{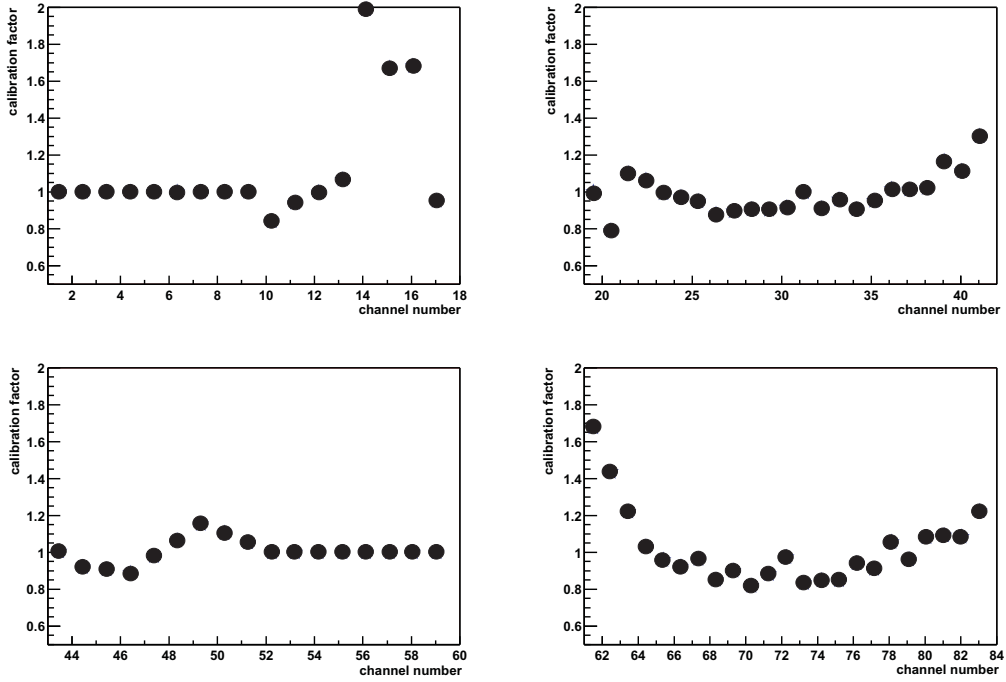}
  \caption{Channel-to-channel calibration factors.}
    \label{fig:KanalAbgleich}
\end{figure}

The second intercalibration iteration has the purpose to correct
for inhomogeneities caused by optical cross talk between adjacent
wavelength shifter bars, which is of the order of $10 $\% \cite{Achim2}.
It is performed after the calibration factors of the first step have
been applied. The procedure is identical to the previous one with the
exception that this time the ``maximum channel'' energy
plus two ``left'' and two ``right'' neighbours are summed up and
filled into the histogram. In this
way, with the iterations carried out as in step one, the ``light 
sharing'' between neighbouring channels due to optical cross talk is 
taken into account. The distribution of the calibration factors
resulting from this procedure, for the first of the two calorimeter 
modules, is shown in Fig. \ref{fig:KanalAbgleich}. The distribution
for the second module is similar \cite{Nix}. The channels with
calibration factors close to 2 suffered from a broken light connection
at one end.

\subsubsection{Determination of the Absolute Energy Scale and
Impact Position Dependent Calibration}

The calibration has to result in correction factors which ensure 
that the energy response for a given particle energy is
the same for all possible impact positions. This was partly done
in the intercalibration steps described above, but due to the
light collection effects mentioned and described in Section 2.3, 
the regions close to the
calorimeter edges are not covered yet. Therefore in the final
step of the calibration procedure the absolute energy scale in 
terms of position dependent calibration factors is determined.

This correction becomes especially important for impact
positions close to the calorimeter edges where parts of the
deposited energy leak out of the active calorimeter volume
into transverse direction, and where scintillator light yield
effects play an important role \cite{GSchmidt}. After
homogenizing the individual channel responses, the absolute
energy scale and the position dependend calibration factors are
determined. 

To homogenize the detector response, the detector
surface is subdivided into segments of $5 \times 5 \,\text{mm}^2$ in
the central region, and $1 \times 1 \,\text{mm}^2$ bins close to the
edges where the response depends strongly on the distance from the
calorimeter edge. The mean energy $\epsilon_{ij}$, averaged over all events
reconstructed in that bin, is calculated. The calibration factor 
$A_{i,j}$ is calculated by normalizing the mean energy in the 
bin $ij$ to the known HERA kinematical peak energy of $27.56\,\text{GeV}$:
\begin{equation}
A_{i,j} = \frac {27.56\,\text{GeV}}{\epsilon_{i,j}}.
\end{equation}

\noindent
The calibration constants have been determined from H1 data for the
channels providing sufficient statistics. The mean value of the constants
is 1.125 with an r.m.s. spread of 0.040 .

\subsubsection{Energy Resolution after Calibration}

\begin{figure}[h]
  \center
      \includegraphics[width=14cm]{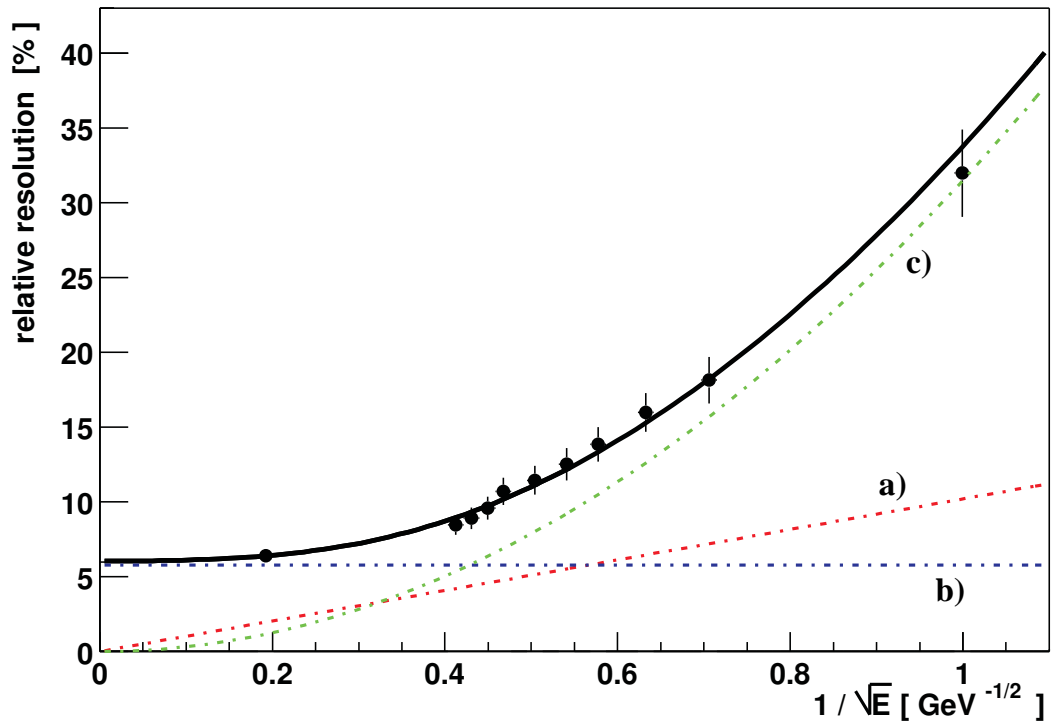}
  \caption{VLQ energy resolution as a function of energy. The dash -
           dotted lines indicate the influence of the contributing terms:
	   a) ``sampling term'', b) ``constant term'', and c) ``noise
	   term''.}
    \label{fig:resolution1}
\end{figure}

For the kinematical peak calibration two data samples were used,
namely from April 1999 and from September - October 1999. The
energy resolution after calibration, as a function of energy and 
from both test beam and kinematical peak data, is displayed 
in Fig. \ref{fig:resolution1}. 
Using the available data at
lower and high energies, an energy resolution parametrized
as

\begin{equation}
\frac{\sigma_E}{E} =\sqrt{\bigg( \frac{(9.16 \pm 4.31)\%}
 {\sqrt{E /\text{GeV}}}\bigg)^2 +  \bigg( (6.04 \pm
0.22)\% \bigg)^2 +
 \bigg(\frac{ (31.94 \pm 3.48)\% }{E/\text{GeV}}\bigg)^2}
\end{equation}

\noindent
is derived. The individual contributions from sampling term, 
constant term, and noise term are indicated in the figure.

When compared to the parametrization of the resolution curve obtained
from test beam data (Eq. 8), a significant increase of the noise term
is observed. This is a consequence of the electronic environment present
in the H1 experimental hall which is very much different from the
situation in the test beam area. The additional point
taken at the HERA electron beam energy constrains the fit at high energy
so that the constant term can be determined with high accuracy. No such 
measurement was available in the test beam. The sampling term is 
consistent with both that obtained from a simulation (Eq. 2) and from
the low energy test beam data.

\section{Performance at HERA}

During its two years of operation in the H1 experiment at HERA, the main
purpose of the VLQ spectrometer was the detection of high energy electrons
and photons in the context of structure function measurements 
\cite{Duprel} and meson
spectroscopy. The performance of the calorimeter is demonstrated best
with purely electromagnetic interactions where no other H1 detector
component is involved, e.g. with wide angle bremsstrahlung events, also
called ``QED Compton (QED-C)'' events \cite{ThKluge}, i.e.
$$  e^-~p~\rightarrow~p~e^-~\gamma .$$

A striking feature of QEDC events is, due to the generally very small
four-momentum transfer to the target proton, the coplanarity of the
final state electron and photon: In the transverse detector plane (``r -
$\phi$ - plane'') photon and electron momentum vector components are back -
to - back. In addition, the sum of photon and final state electron energy
must equal the initial electron beam energy.

\begin{figure}[h]
\begin{center}
       \includegraphics[width=10cm]{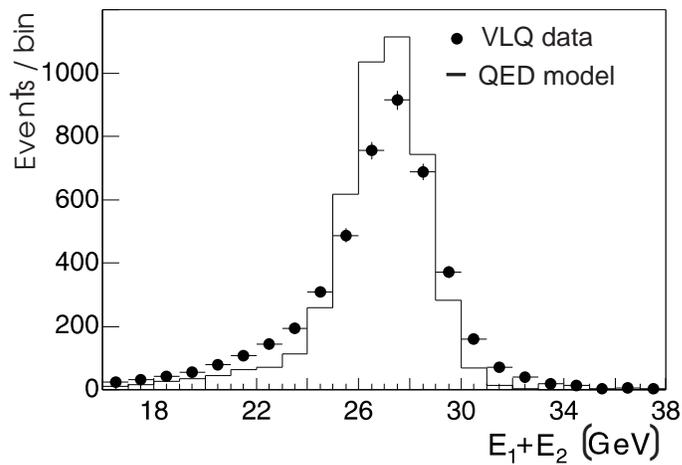}
 \begin{quote}
  \caption{Sum of photon- and electron energies for QED - Compton events
           measured in the VLQ calorimeter. Comparison of data and  
           QED simulation.}
    \label{fig:QEDC1}
 \end{quote}
\end{center}
\end{figure}

\begin{figure}[h]
 \begin{center}
      \includegraphics[width=10cm]{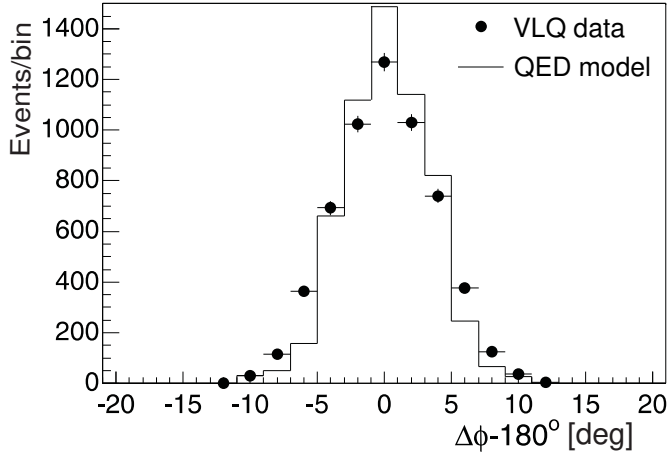}
 \begin{quote}
  \caption{Azimuth angular difference of photon and electron impact points
           in the VLQ calorimeter, for QED Compton events. Comparison of
           data and QED simulation.}
    \label{fig:QEDC2}
  \end{quote}
\end{center}
\end{figure}

The VLQ calorimeter is ideally suited to test these properties (or, to
profit from them because they can be used for calibration and
alignment purposes as described in \cite{ThKluge}). The performance of
the calorimeter is demonstrated in Figs. \ref{fig:QEDC1} and 
\ref{fig:QEDC2}, taken from \cite{ThKluge}. Fig. \ref{fig:QEDC1} shows the
sum of final state electron and photon energies, together with the 
result of a simulation, and Fig. \ref{fig:QEDC2} shows the azimuth angular 
difference, shifted by $180^\circ$, of photon and electron impact points, 
again compared to a simulation. Apparently, the calorimeter behaved as
expected.

In the context of an analysis searching for single $\pi^0$'s produced
in the H1 backward region \cite{oddpom}, two - photon pairs were
collected and investigated for a signal, where either both photons
had to be detected in the VLQ calorimeter, or one photon in the VLQ
and the other one in the SpaCal backward calorimeter.
The low cross section, in combination
with special kinematic restrictions, did not leave a sizeable signal
for the VLQ - VLQ case alone.

\begin{figure}[h]
  \begin{center}
     \includegraphics[width=10cm]{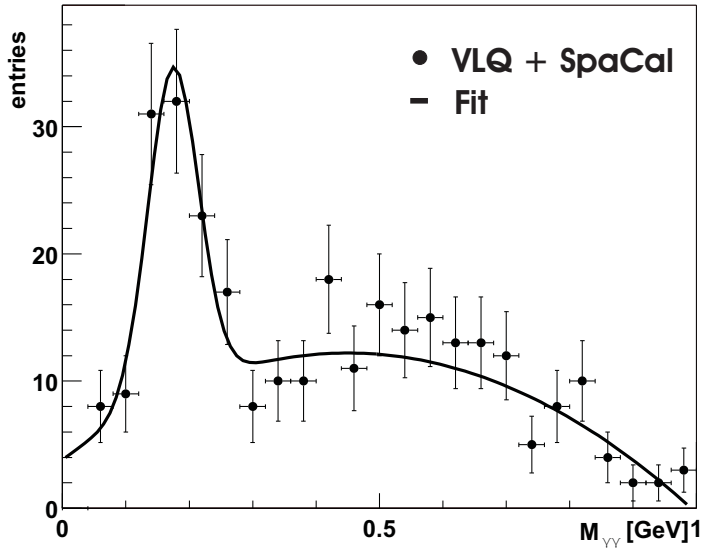}
  \begin{quote}
  \caption{Two - photon invariant mass distributions from photoproduction
           events where one photon is measured in the VLQ, and one in 
           the SpaCal.}
    \label{fig:twogam}
    \end{quote}
    \end{center}
\end{figure}

A clear $\pi^0$ signal was observed, however, if the combination
VLQ - SpaCal is added, see Fig. \ref{fig:twogam} where a fit of a
Gaussian function to the $\pi^0$ peak plus a polynomial for the 
background is superimposed. The $\pi^0$ peak
is shifted towards a mass larger than the PDG value, and appears
broader than expected for both SpaCal and VLQ mass resolutions
\cite{Tobi}. There are several reasons responsible for these
effects: First, the position of the actual interaction vertex is not
known because there are no charged particle tracks in these events,
and four-momenta have to be reconstructed using the nominal vertex
position.
Second, both the SpaCal and VLQ calibrations were performed with the
help of the ``Kinematical Peak'', i.e. with electrons of an energy
close to the incident electron beam energy (see Section 4.2) while
photons from $\pi^0$ decays have much smaller energies. Third, 
no dedicated intercalibration was performed between the SpaCal and VLQ
calorimeters. In spite of these limitations, the capability of the VLQ
to contribute to the reconstruction of $\pi^0$'s from their decays to
two photons is convincingly demonstrated.

\section{Conclusions}

The Very Low $\text{Q}^2$ calorimeter of H1, a highly compact device 
designed to measure high energetic electrons and photons in the
electron beam downstream region, was succcessfully operated in the
1999 and 2000 data taking periods. The performance in the experiment 
was very close to that expected from design and simulations. The VLQ
calorimeter contributed significantly to measurements of the proton 
structure function $F_2$ in phase space regions previously unaccessible 
for H1, and opened a new kinematical regime for the spectroscopy of 
mesons decaying to purely photonic final states.

\section*{Acknowledgements}

We gratefully acknowledge the assistence and skill of the electronic 
and mechanical workshops of the Kirchhoff - Institut f\"ur Physik of 
the Universit\"at Heidelberg in designing and constructing the 
calorimeter. We are indebted to the DESY Hallendienst for providing
the test beam facilities. It is a pleasure to thank K. Gadow (DESY)
for his enthusiastic and untiring support during the installation 
and data taking phases of the VLQ spectrometer. Financial support from
the Bundesministerium f\"ur Forschung und Technologie, FRG, 
under contract numbers 6HD27I, 5H11VHB/5, the Deutsche Forschungsgemeinschaft, 
and VEGA SR, grant no. 2/1169/2001, is gratefully acknowledged.

\end{document}